\documentstyle[12pt,epsfig]{article}
\begin{document}

\begin {center}
{\Large The $\pi \pi$ mass spectrum in  $Y(4260) \to \pi \pi J/\psi $.}
\vskip 3mm

{D.V.~Bugg\footnote{email address: D.Bugg@rl.ac.uk}}   \\
{Queen Mary, University of London, London E1\,4NS, UK}
\end {center}
\vskip 2.5mm

\begin{abstract}
Three ways of fitting the $\pi ^+\pi ^-$ mass spectrum in $Y(4260)$
decays are studied.
Data presented recently by Belle cannot be fitted by the $\pi \pi$
S-wave intensity for elastic scattering.
They can be fitted by adding a rather arbitrary destructive
interference with the $\sigma$ pole term.
A better fit may be obtained with the decay sequence
$Y \to \pi H$, $H \to \pi J/\psi $, where $H$ is a $J^{PC}=1^{--}$
$c\bar c q \bar q$ state peaking at 4.0 GeV with a width
$\sim 280$ MeV.
A third possibility, involving a triangle diagram due to
$Y(4260) \to DD_1(2420)$, $D_1 \to D^*(2007)\pi$,
$DD^* \to \pi J/\psi$ fails to fit the data.
The first and second possibilities (or a combination) could be resolved
by analysis of the $Y(4260)$ Dalitz plot and $\pi J/\psi$ mass
projection, which are not presently publicly available.

\vspace{5mm}

\end{abstract}

There has been considerable interest in the $Y(4260)$ discovered
by BaBar [1].
It has been confirmed by Cleo [2], by further observations by BaBar
[3,4] and Cleo [5] and most recently by Belle [6].
Its clear observation in decays to $\pi \pi J/\psi $ has led
to suggestions that it may be related to the $c\bar c g$ hybrids
predicted in this general mass range [7-10].
It is observed in the radiative return process, where it is produced
through $e^+e^- \to \gamma$; it therefore has $J^{PC} = 1^{--}$.
Cleo observe $Y(4260) \to \pi ^0 \pi ^0 J/\psi $ [2] with
a rate consistent with half that of $\pi ^+\pi ^- J/\psi$,
albeit with just 8 clean events.
This result strongly supports $I = 0$ for $Y(4260)$.

The latest Belle data [6] reveal a curious $\pi \pi$ mass spectrum.
The objective here is to examine ways of fitting this spectrum.
Three classes of decay scheme for $Y(4260)$ are investigated.

The simplest hypothesis is $Y \to (\pi \pi )_S J/\psi $, where
$(\pi \pi)_S$ stands for the S-wave as observed in elastic scattering.
The full curve of Fig. 1(a) compares data points with the spectrum
predicted from the latest parametrisation [11]; a brief subroutine
for this parametrisation is available from the author.
In making this prediction, the line-shape of $Y(4260)$ has been
folded with the phase space for $Y \to (\pi \pi )_S J/\psi$
using the Breit-Wigner parameters determined by Belle.

%Fig. 1
\begin {figure}  [htp]
\begin {center}
\vskip -4.0cm
\epsfig {file=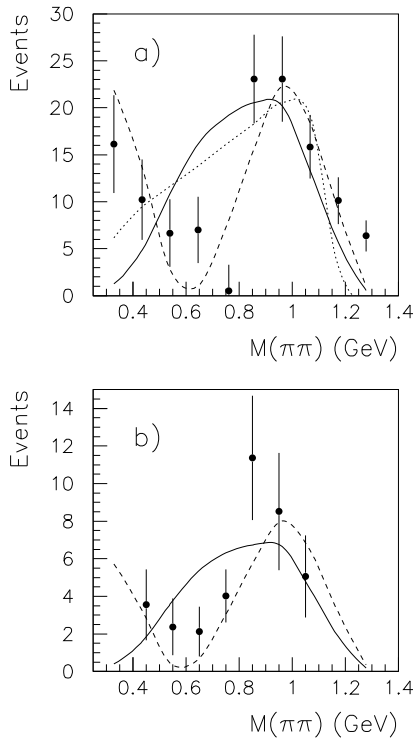,width=7.5cm}\
\vskip -4mm
\caption {(a) The $\pi \pi$ mass spectrum from Belle data for
$Y(4260) \to \pi ^+\pi ^- J/\psi $ (data points), compared with
(i) production of $J/\psi$ with the elastic $\pi \pi $ S-wave (full
curve), (ii) including destructive interference with the $\sigma$
pole using eqn. (2) (dashed curve), (iii) with the triangle
diagram of Fig. 3 (dotted); (b) the corresponding comparison with
Cleo data.}
\end {center}
\end {figure}

The predicted peak at $\sim 900$ MeV is  too wide, fails to
reproduce the low point at 750 MeV and does not account for the low
mass peak.
A minor detail is that none of the fits reported here reproduce
accurately the highest mass point at $M(\pi \pi) \sim 1275$ MeV. This
mass is outside the range accessible for Belle's central $Y$ mass, 4295
MeV; it is possible that events in this bin arise from spreading of the
high mass $\pi \pi$ peak due to mass resolution. This bin is therefore
omitted from all fits, and the curve is normalised to the 113 events in
the remaining 9 bins. The resulting fit has $\chi^2 = 69$ for 9 bins
and 8 degrees of freedom.

In elastic scattering, the $\pi \pi$ amplitude is
%Eqn. (1)
\begin {equation}
f_S(\pi \pi) = \frac {N(s)}{D(s)} =
\frac {M\Gamma (s_{12})}{M^2 - s_{12} - iM\Gamma (s_{12})}
\end {equation}
where $\Gamma (s)$ is close to linear in $s$, with a zero at
the Adler point $s = s_A = m^2_\pi /2$.
[The convention adopted here is that the $J/\psi$ is taken as
particle 3, hence $s_{12}$ is the invariant mass of $\pi \pi$].
The $\pi \pi$ phase shift goes through $90^\circ$ at $M \sim 950$
MeV and the intensity of $\pi \pi$ elastic scattering peaks there.
However, the pole in the amplitude lies much lower, at $\sim 470$
MeV, because of the strong $s$-dependence of the width.
In some production processes, e.g. $J/\psi \to \omega \pi \pi$ [12]
and $D \to 3\pi$ [13], the production amplitude is instead well
fitted by the same Breit-Wigner denominator $D(s)$ as eqn. (1), but
with a constant numerator $N(s)$. The elastic amplitude seems to give
a good fit to processes involving low momentum transfers, e.g. decays
$\psi (2S) \to \pi \pi J/\psi $ [14]and $\phi \to \gamma \pi ^0\pi ^0$
[11].
The production processes of Refs. [12] and [13] involve larger
momentum transfers, hence more distant singularities.
The decay $Y(4260) \to \pi \pi J/\psi$ is closer to the former case.

An adequate fit may be obtained using destructive
interference between Eqn. (1) and the $\sigma $ pole by altering the
numerator $N(s)$ to
\begin {equation}
f_S(\pi \pi) =
\frac {(s - s_A) - G(M^2 - s_A)}{D(s)},
\end {equation}
with the same Breit-Wigner denominator $D(s)$ as in eqn. (1).
The resulting fit is shown by the dashed curve in Fig. 1,
with $G = 0.37$, $\chi^2 = 17.8 $ for 9 bins (7 fitted parameters).
This fit seems somewhat artificial, since the numerator has no
obvious explanation; however, it cannot presently be ruled out.
Making $G$ complex does not improve the fit significantly.

It seems unlikely that $f_0(980)$ is responsible for the
peak at 950 MeV in Fig. 1, because the $f_0$ is  too narrow.
The latest BES data on $J/\psi \to \phi \pi \pi$ and $\phi KK$ [15]
determine the full-width at half maximum of $f_0(980) \to \pi \pi$
to be $34 \pm 8$ MeV.
Belle do not quote their $\pi \pi$ mass resolution, but do quote
the width of $Y(4260)$ as $133 \pm 26 ^{+13}_{-6}$ MeV.
The mass resolution in both cases should be comparable.
In any case, the $\pi \pi $ mass distribution cannot be fitted by
$f_0(980)$ and the $\sigma$ pole: the contribution from the $\sigma$
pole is too broad if it interferes with only the very narrow $f_0(980)$.
Cleo observe 2 events for the decay to $K^+K^-J/\psi$ [2].
These could arise from the known amplitude for $\sigma \to KK$
or from a small $f_0(980)J/\psi$ contribution.

Fig. 1(b) compares the same fits with the 37 events of Cleo [2].
Although statistics are lower, the peak in data again looks narrower
than the full curve; there is also some tentative indication of a rise
at low $\pi \pi$ mass.
Cleo correct for background from sidebins around the $J/\psi$, but
do not correct for possible background under the $Y(4260)$ from other
$\psi$ states. Judging from the mass range 4.0 to 4.2 GeV, this
background could be $15\%$.

BaBar [1] show a $\pi \pi$ mass spectrum which does not show any rise
at low $\pi \pi$ mass.
It has been obtained by summing the  $Y(4260)$ signal over the whole
mass range and then making a background subtraction.
The resulting signal/background ratio is $\sim 1:1$.
In this respect, Belle's strategy of selecting events in the mass
range 4.2 to 4.4 GeV, i.e. 1.5 times the full width, gives a better
signal/background ratio for the purpose of viewing the $\pi \pi$
mass spectrum. Some such weighting of data points near the
peak of $Y(4260)$ is desirable.

Let us turn to a second hypothesis: sequential
decays of the form $Y \to \pi H$, $H \to \pi J/\psi $.
If $Y$ has $I = 0$, as Cleo data suggest, the $H$ has $I = 1$ and
could be a 4-quark state.
Two cases need consideration.
When the orbital angular momentum $L$ in both steps of
the sequence is zero, $H$ has $J^{PC} = 1^{+-}$.
If $L =1$, $H$ has $J^{PC} = 1^{--}$, $0^{--}$ or $2^{--}$.

With $L = 0$ in both steps, the predicted $\pi \pi$ mass spectrum
is close to the full curve of Fig. 1 and hence fails to fit
the data.
However, with $L = 1$, a good fit emerges naturally,
as shown on Fig. 2(a).
This result depends on simple kinematics.
Consider the case $Y_{123} \to \pi _1H_{23}$, $H_{23} \to \pi_2
(J/\psi)_3$, where numbers label the particles. If $L = 1$ in the
second step, there is a factor $\cos \theta _{2}$ in the amplitude;
here $\theta _2$ is the angle of $\pi _2$ in the $23$ rest frame with
respect to the recoil direction of particle 1.
For fixed $s_{23}$, $s_{12}$ varies linearly with $\cos \theta _2$
from one side of the Dalitz plot to the other, leading to a zero
halfway between left- and right-hand edges of the plot.
The mass of $H$ determines the value of $s_{12} = m^2_{12}$ at which
the zero occurs, via the boundaries of the Dalitz plot.
The width of $H$ controls the sharpness of the dip in the mass
spectrum near 750 MeV.

%Fig. 2
\begin {figure}  [t]
\begin {center}
\vskip -40mm
\epsfig{file=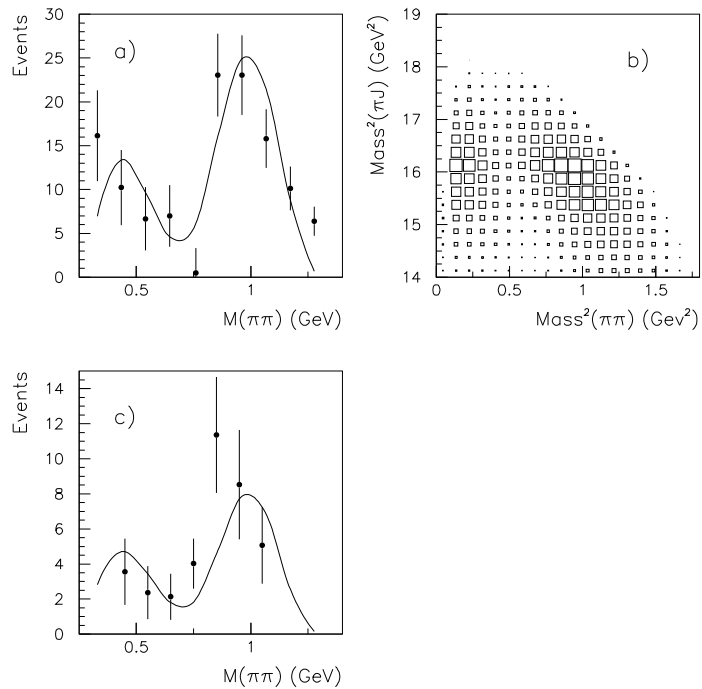,width=10cm}\
\vskip -5mm
\caption{(a) Fit to the Belle $\pi \pi$ mass spectrum for $L=1$
decays via an intermediate $H$ with
$M = 4080$ MeV, $\Gamma = 280$ MeV; (b) the upper half of
the Dalitz plot; (c) fit to Cleo data.}
\end {center}
\end {figure}

Fig. 2(b) shows the upper half of the Dalitz plot for this hypothesis.
The zero due to the factor $\cos \theta _2$ is clearly visible.
Fig. 2(c) shows the fit to Cleo data.
The fitted mass and width are almost linearly correlated because of the
$L=1$ centrifugal barrier.
The optimum fit has a mass of 4.08 GeV with $\Gamma = 0.275$ GeV
and $\chi^2 = 15.1$ for 2 fitted parameters, seven degrees of freedom.
Fits poorer in $\chi^2$ by 1 are obtained with $M = 4.04$ GeV,
$\Gamma = 0.225$ GeV or $M = 4.16$ GeV, $\Gamma = 0.330$ GeV.
Because of the centrifugal barrier for production, the peak
intensity is just below 4 GeV for all solutions.
A $Y$ mass at the lower end of the range is most plausible, in order to
give a reasonable separation in masses of $Y$ and $H$.

In detail, the amplitude for production of $H_{23}$
alone is
\begin {equation}
f \propto f_Y(s_{123}) f_H(s_{23})B_1(p_1)B_1(p_2)\cos \theta _2,
\end {equation}
where $f_Y$ is the Breit-Wigner amplitude of $Y$ as a function of
$s_{123}$,
$f_H$ is the Breit-Wigner amplitude for $H$,
$B_1(p_1)$ is the $L=1$ centrifugal barrier factor as a function of
the momentum of $\pi_1$ in the rest frame of $Y$, and $B_1(p_2)$ is
the centrifugal barrier factor as a function of the momentum of
$\pi_2$ in the rest frame of $H_{23}$.
The centrifugal barrier is taken as
$B_1(p) = p/\sqrt {p^2 + p_0^2}$ where $p_0 = 0.25$ GeV/c.
The centrigual barrier cuts off the intensity near the top of Fig. 2(b).
The amplitude of eqn. (3) needs to be symmetrised between particles 1
and 2.
The usual weighting according to phase space $ds_{12}ds_{23}$ is
also applied.

None of the experimental groups has presented data
as a function of $m^2(\pi J/\psi )$.
Even with the current low statistics, structure like that
of Fig. 2(b) should be clearly visible.
It gives rise to a strong peaking at two corners of the Dalitz
plot and a peak in the $\pi J/\psi$ spectrum near 4.0 GeV.
The situation could however be confused by interferences
between amplitudes of the two forms considered so far.
Furthermore, interference between $Y(4260)$ and underlying
background from $\psi (4160)$ and $\psi (4415)$ is possible.
Indeed, van Beveren and Rupp have argued against the existence
of $Y(4260)$ as a separate resonance [16];
they interpret the structure in this mass range as arising
from thresholds and interferences in the $s$-dependence of
$\psi (4040)$ and $\psi (4160)$ amplitudes.

%Fig. 3
\begin {figure}  [htp]
\begin {center}
\vskip -43mm
\epsfig{file=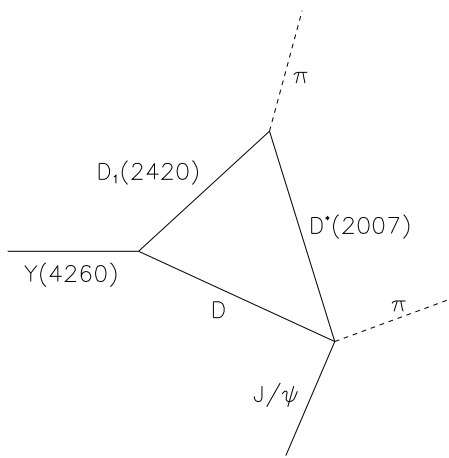,width=12cm}\
\vskip -28mm
\caption{The triangle diagram for $Y \to D D_1(2420)$,
$D_1(2420) \to \pi D^*(2007)$, $D D^*(2007) \to \pi J/\psi $.}
\end {center}
\end {figure}

A third hypothesis will now be considered.
The $Y(4260)$ lies close to the threshold for $D(1865)D_1(2420)$,
particularly for the resonance mass quoted by Belle, 4295 MeV.
A possible mechanism for production of the $\pi \pi J/\psi $
system is shown in Fig. 3.
In this so-called triangle diagram, the $D_1$ decays (via a D-wave)
to $\pi D^*(2007)$.
This long-lived resonance can de-excite from the spectator $D$
producing $\pi J/\psi $.
A simulation of this process gives the dotted curve of Fig. 1(a),
and fails to fit the Belle data.
Even if interference with the $(\pi \pi )_S J/\psi$ amplitude is
included with a fitted complex coupling constant, nothing resembling
the Belle mass spectrum can be reproduced.
The essential feature of the
process in Fig. 3 is that the $D$ and $D_1$ are produced almost at rest
in the $Y$ rest frame. The pion from  $D_1(2420)$ decay has a momentum
close to 360 MeV/c in the $D$ rest frame and hence in the $Y$ rest
frame. This makes a clear signature for this process.

In summary, the $\pi \pi$ mass spectrum observed by Belle in
$Y(4260)$ decays is inconsistent with that expected for a
final state $(\pi \pi )_S J/\psi$ unless the numerator of the
amplitude gives rise to destructive interference.
The simplest, but speculative, hypothesis is the sequential decay
to $J/\psi$ via an intermediate $J^{PC} = 1^{--}$ 4-quark state.
This possibility fits the data unaided;
on the other hand, there is no obvious reason why the $Y(4260)$ alone
should prefer decays to $\pi \pi J/\psi$.
It should be simple to verify or reject this hypothesis in its
simplest form, although interferences with the first mechanism
could complicate the situation.

\section*{Acknowledgement}
This work began during a stimulating visit to the
Centro de F\'{\i}sica das Interac\c{c}\~{o}es Fundamentais of the
Instituto Superior T\'{e}cnico in Lisbon.
I wish to thank Profs. E. van Beveren and G. Rupp for their
hospitality and extensive discussions on this and related
topics.
I am grateful to Dr. F.-K. Guo for correcting a mistake in the
original manuscript.

\end {document}